\documentclass[twocolumn,british,useAMS,usenatbib]{mn2e}
\usepackage{mathptmx}

\usepackage[T1]{fontenc}
\usepackage[utf8]{inputenc}
\usepackage{babel}
\usepackage{color}
\usepackage{textcomp}
\usepackage{amsmath}
\usepackage{amssymb}
\usepackage{graphicx}
\usepackage[authoryear]{natbib}
\usepackage{subscript}

\usepackage{mathrsfs}



\begin{document}

\title{On the accuracy of close stellar approaches determination}

\author[Piotr A. Dybczy{\'{n}}ski and Filip Berski]{Piotr A. Dybczy{\'{n}}ski$^{1}$\thanks{E-mail: dybol@amu.edu.pl} and Filip Berski$^{1}$\thanks{E-mail: filip.berski@amu.edu.pl}\\
$^1$Astronomical Observatory Institute, Faculty of Physics,
A.Mickiewicz Univ., S\l{}oneczna 36, 60-286 Pozna\'{n}, Poland;}
\date{Accepted xxxx. Received xxxx; in original form xxxx}

\maketitle

\begin{abstract}
The aim of this paper is to demonstrate the accuracy of our knowledge
of close stellar passage distances in the pre-GAIA era. We used the
most precise astrometric and kinematic data available at the moment
and prepared a list of 40 stars nominally passing (in the past or future)
closer than 2 pc from the Sun. We used a full gravitational potential
of the Galaxy to calculate the motion of the Sun and a star from their
current positions to the proximity epoch. For this calculations we
used a numerical integration in rectangular, Galactocentric coordinates.
We showed that in many cases the numerical integration of the star
motion gives significantly different results than popular rectilinear
approximation. We found several new stellar candidates for close visitors
in past or in future.

We used a covariance matrices of the astrometric data for each star
to estimate the accuracy of the obtained proximity distance and epoch.
To this aim we used a Monte Carlo method, replaced each star with
10\,000 of its clones and studied the distribution of their individual
close passages near the Sun. We showed that for contemporary close
neighbours the precision is quite good but for more distant stars
it strongly depends on the quality of astrometric and kinematic data.
Several examples are discussed in detail, among them the case of HIP~14473.
For this star we obtained the nominal proximity distance as small
as 0.22 pc 3.78 Myr ago. However there exist strong need for more
precise astrometry of this star since the proximity point uncertainty
is unacceptably large. 
\end{abstract}

\section{Introduction}

\label{sec:Introduction}

Soon after the Oort\citeyearpar{oort:1950} published a paper on the
existence of the comet cloud surrounding the Sun, many authors started
to investigate the influence of real nearby stars on cometary orbits.
One of the earliest complete (at that moment) search for stellar passages
near the Sun was published by \citet{makover:1964}. He listed all
past stellar proximity epochs and distances based on the first edition
of the Gliese Catalogue \citeyearpar{gliese:1957} and obtained with
a rectilinear approximation. In recent twenty five years several papers
were published in this field, among others \citealp{matthews:1994,mullari-o:1996,dyb-kan:1999,garcia-sanchez:1999,garcia-sanchez:2001,dyb-hab3:2006,bobylev:2010,bobylev:2010-1,jimenez_et_al:2011}.
The main reason for a continuous interest in determining close star
passages is the progress in obtaining stellar data, especially proper
motions, parallaxes and radial velocities. Additionally, investigations
of long period comets past and future dynamics are more and more detailed
(see for example \citealp{kroli-dyb:2010,kroli-dyb:2013,dyb-kroli:2011,dyb-kroli:2015,kroli-dyb:2012,krolikowska:2014}).
As a result best possible knowledge on potential stellar perturbers
of such motion is of great interest \citep{fouchard-r-f-v:2011}.

With the publication of the first Hipparcos catalogue \citeyearpar[hereafter HIP1]{hipparcos-cat:1997}
the number of stars with known parallaxes has increased significantly.
Hipparcos mission \citep{hipparcos} was a milestone in our last decades
advance in gathering data on the spatial distribution of stars in
the solar neighbourhood. This mission gave us precise positions, parallaxes
and proper motions for \textasciitilde{}120 thousands of stars from
carefully and in advance selected list, forming the Hipparcos Input
Catalogue \citep{HIC-1:1992}. Hipparcos catalogue is complete up
to $\sim25$ pc for all stars brighter than $M_{\textrm{V}}\cong9$
mag \citep{jehreiss_wielen:1997} and for the massive stars in this
population the completeness reaches a much larger distance. There
are two main reasons for which it is difficult to obtain precise information
on the spatial velocity for all Hipparcos stars. First - the Hipparcos
proper motions, while of great formal precision, are based on relatively
short time interval, therefore for a large number of stars they differ
from the mean, secular proper motions \citep{wielen-etal:1999a}.
Several attempts were made to combine Hipparcos proper motions with
those from a ground based observations, see for example \citep{wielen-etal:1999b,hoogerwerf-blaauw:2000,wielen:2001}.
The most fruit-full attempt was the construction of Tycho-2 catalogue
\citep{tycho2-cat:2000,tycho2-constr:2000} however combining Tycho-2
proper motions with parallaxes from HIP-1 introduces some inconsistency
in treating astrometric data. Recently a new reduction of Hipparcos
raw data were performed and a second, significantly improved version
of the Hipparcos catalogue (HIP2, \citealp{vanleeuwen:2007,vanleeuwen:2011}
) have been published. It improves significantly precision of both
proper motions and parallaxes.

As it concerns radial velocity, several large projects (see for example
\citealp{grenier-etal:1999b,grenier-etal:1999a,nidever-etal:2002,nordstrom:2004})
increased the number available measurements and a large Pulkovo compilation
of radial velocities of 35 495 Hipparcos stars was published by \citet{Gontcharov:2006}.
Using HIP2 catalogue, Pulkovo compilation of radial velocities and
several other sources of astrophysical parameters \citet{anderson_francis:2011}
published an extended Hipparcos compilation of stellar data, known
as the XHIP catalogue.

The second reason for investigating stellar passages is our increasing
ability to calculate the stellar path relative to the Sun with increasing
accuracy, taking into account the Galactic gravity field. Using such
advanced methods we should additionally ask about the accuracy of
the results and their dependency on the used data.
 Authors of several previous papers also estimated the accuracy of their results
in more or less approximate manner. For example \citet{garcia-sanchez:2001}
simply used the root of a sum of squared error contributions from
two components of proper motions, parallaxes and radial velocities.
The Monte Carlo method of estimating the accuracy of the distance
of a stellar close passage near the Sun was first used by \citet{mullari-o:1996}
but in a simplified manner (for example with assumed 3 km\,s$^{-1}$
error for all radial velocities). Similar attempt was performed by
\citet{bobylev:2010,bobylev:2010-1} and recently by \citet{jimenez_et_al:2011}
but in all cases only formal errors were taken into account, ignoring
mutual correlations between astrometric parameters.

In the present work we used astrometric parameters taken directly from
the HIP2 catalogue instead of XHIP mainly because HIP2 presents also
a covariance matrix which is necessary to apply the advanced method
for the accuracy assessment. We used radial velocities from the XHIP
catalogue. For several selected stars we presented comparison of the
results based on astrometry and radial velocities from some other
sources.

This paper provides the most up to date information on the closest
stellar approaches to the Sun, obtained from latest astrometric data
and radial velocities augmented with the more elaborated Monte Carlo
assessment of the accuracy of the obtained minimal distances and their
epochs, based on full covariance matrices included in the HIP2 catalogue.
Of course we expect significant improvements from the Gaia mission
therefore our computer codes are fully prepared to include new data.

In the next section we describe methods of our calculations. Section
\ref{sec:Results-for-selected} consists of the results obtained for
selected stars. In section \ref{sec:Just-in-the} we present a discussion
with a very recent paper by \citet[hereafter B-J]{bailer-jones:2014},
indpendently prepared and made available to the public when this work
was finished. In the last section some conclusions and prospects are
drawn.

\section{Methods of calculations}

\label{sec:Methods-of-calculations}

\subsection{Units, definitions, reference frames}

\label{sub:Units-definitions}

To achieve the aim of this work we have to study stars (including
the Sun) motion under the gravitational influence of the Galaxy. This
can be performed only for stars with the full 6D data available, typically
expressed in the equatorial frame, i.e. right ascension $\alpha$,
declination $\delta$, parallax $\pi$, proper motions $\mu_{\alpha*}$,
$\mu_{\delta}$ (a star subscript denotes the multiplication by $\cos\delta$
) and radial velocity $v_{r}$. The starting data for a numerical
integration, namely position and velocity components in the Galactocentric
frame are to be obtained as follows. We first calculate the heliocentric
distance of a star from the formula:

\begin{equation}
r_{h}=\frac{1000}{\pi}\label{eq:distance}
\end{equation}
where the parallax $\pi$ is expressed in miliarcseconds {[}mas{]}
what gives $r_{h}$ in parsecs. According to comments expressed by
\citet[chapter 3, page 86 ]{hip2_book:2007} we resisted here to add
the so called Lutz-Kelker bias correction, proposed by \citet{anderson_francis:2011}.

Next, heliocentric, equatorial rectangular coordinates of a star (in
parsecs) are given by:

\begin{eqnarray}
 &  & x_{h}=r_{h}\cdot\cos\alpha\cos\delta\nonumber \\
 &  & y_{h}=r_{h}\cdot\sin\alpha\cos\delta\label{eq:xyz_helio}\\
 &  & z_{h}=r_{h}\cdot\sin\delta\nonumber 
\end{eqnarray}

If proper motions are given in mas/yr and radial velocity is expressed
in km\,s$^{-1}$ than the heliocentric, equatorial rectangular
velocity components (in pc/Myr) might be calculated as follows:

\begin{eqnarray*}
 &  & v_{1}=s\cdot r_{h}\cdot\mu_{\alpha*}/\cos\delta\\
 &  & v_{2}=s\cdot r_{h}\cdot\mu_{\delta}\\
 &  & v_{3}=k\cdot v_{r}\\
 &  & \dot{x}_{h}=-\cos\delta\sin\alpha\cdot v_{1}-\sin\delta\cos\alpha\cdot v_{2}+\cos\delta\cos\alpha\cdot v_{3}\\
 &  & \dot{y}_{h}=\cos\delta\cos\alpha\cdot v_{1}-\sin\delta\sin\alpha\cdot v_{2}+\cos\delta\sin\alpha\cdot v_{3}\\
 &  & \dot{z}_{h}=\cos\delta\cdot v_{2}+\sin\delta\cdot v_{3}
\end{eqnarray*}

where $s=0.0048481368$ is a coefficient for angular units conversion
and $k=1.022689369$ is a coefficient for velocity units conversion.
Throughout this paper we use the following units: parsec {[}pc{]}
as the distance unit, solar mass {[}M$_{\sun}${]} as the mass unit
and million of years {[}Myr{]} as the time unit. The constant of gravity
expressed in these units is: G$=4.498297316\times10^{-3}$. 

Reorientation of the equatorial frame into Galactic one (still heliocentric)
involves three rotations: 
\[
\mathbf{r}_{\textrm{h,gal}}=R_{z}(90\degr-\theta)R_{x}(90\degr-\delta_{o})R_{z}(90\degr+\alpha_{o})\mathbf{r}_{\textrm{h,equ}}
\]
where $\theta=122\fdg93191857$, $\alpha_{o}=12^{h}51^{m}26\fs27549$
and $\delta_{o}=27\degr07\arcmin42\farcs7043$ are the positional
angle and north Galactic pole equatorial coordinates defining the
Galactic frame orientation while $R_{x}$ and $R_{z}$ denote the
rotation with respect to $OX$ and $OZ$ axes respectively. See \citet{liu_et_al:2011}
for a recent discussion on the Galactic frame orientation.

To move the origin from the Sun to the Galactic centre it is necessary
to add the Galactic position and velocity of the Sun:

\[
\mathbf{R}=\mathbf{r}_{\textrm{h,gal}}+\mathbf{R}_{\sun},\quad\mathbf{\dot{R}}=\dot{\mathbf{r}}_{\textrm{h,gal}}+\mathbf{\dot{R}}_{\sun},
\]

The values for the solar position $\mathbf{R_{\sun}}=(x_{\sun},y_{\sun},z_{\sun})$
and velocity $\mathbf{\dot{R}}_{\sun}=(u,v,w)$ components should
be chosen in accordance with the adopted Galactic potential model.
In our calculation we use $\mathbf{R_{\sun}}=(x_{\sun},y_{\sun},z_{\sun})=(-8400,0,17)$
in pc. Since the vertical position of the Sun with respect to the
Galactic disk plane is still uncertain we decided to follow arguments
of \citet{joshi:2007} and adopt $z_{\sun}=$+17 pc. The reader should
be warned that during this translation we kept the orientation of
all axes, so in a Galactocentric frame the $OX$ axis is directed
opposite to the Sun direction, the $OY$ is directed in accordance
to the Sun rotation around the Galactic centre and the $OZ$ axis
is directed to the north Galactic pole, what makes this system a right-handed
one.

As it concerns the Sun peculiar velocity, we use values proposed by
\citet{irrgang_et_al:2013}, what after adding the Sun Galactic rotation
velocity resulting from the adopted potential model gives:
\[
\mathbf{\dot{R}}_{\sun}=(u,v,w)=(+11.1,+254.24,+7.25)\; km\,s^{-1}
\]

\[
=(+11.352,+260.011,+7.41)\; pc/Myr.
\]

\subsection{Drawing a stellar clone}

\label{sub:Drawing-a-stellar}

In order to estimate the uncertainty of the close stellar passage
parameters we replace each considered star with a large number (typically
10\,000) of its clones, drawn from a multivariate normal distribution.
This is possible because in HIP2 catalogue \citet{vanleeuwen:2007,vanleeuwen:2011}
included full information on the covariance matrix of astrometric parameters. Such a procedure possesses an evident
superiority over the individual, independent random drawing of each
parameter (which was used for example by \citet{bobylev:2010-1,bobylev:2010,jimenez_et_al:2011})
what ignores their obvious mutual dependence. We demonstrate this
at the end of this section. There are no radial velocities in HIP2
so we have to draw radial velocity independently using data from XHIP
catalogue. Unfortunately for some stars the formal error is unknown
and the authors insert 999 km\,s$^{-1}$ in catalogue. One of such  stars is
HIP 21539 for which we adopted 30 km\,s$^{-1}$ as formal error. We used this
value because this radial velocity measurement \citep{barbier-brossat:1994}
most probably comes from observation with an obiective prism which
have the precision of this order. 

To generate random vectors $\mathbf{x}$ from a multivariate normal
distribution $\mathcal{N}_{n}(\mathbf{m},\,\Sigma)$ with a given
vector of means $\mathbf{m}$ an a given covariance matrix $\Sigma$
one can proceed as follows:
\begin{enumerate}
\item Decompose the given covariance matrix (e.g. with Cholesky method)
to obtain the matrix $\mathbf{G}$ such, that $\Sigma=\mathbf{G}\mathbf{G}^{T}$. 
\item Compose a vector $\mathbf{y}$ of $n$ independent random scalars
$y_{i}$ from a standard normal distribution: $y_{i}\sim\mathscr{\mathcal{N}}(0,1)$. 
\item Obtain vector $\mathbf{x}$ from: $\mathbf{x}=\mathbf{m}+\mathbf{G}\mathbf{y}$. 
\end{enumerate}
Here are some details on applying the above recipe for a HIP2 star
cloning. In what follows we assume the 5 parameter solution but the
analogous approach can be applied to stars with the 7- or 9-parameters
solutions. For each star in HIP2 catalogue given is an upper triangular
'weight matrix' $\mathbf{U}$. According to the catalogue description
it should be related to the covariance matrix $\Sigma$ by a simple
relation:

\begin{equation}
\Sigma^{-1}=\mathbf{U}^{T}\mathbf{U}\label{eq:z_hip2}
\end{equation}

However this is not exactly true (van Leeuwen, 2013, personal communication).
To obtain matrix $\mathbf{U}$ from catalogue data it is necessary
to re-scale its diagonal elements ($u_{1}$,$u_{3}$,$u_{6}$,$u_{10}$
and $u_{15}$ from catalogue) multiplying each of them by a factor
of $(\sigma_{0}/\Delta x_{i})$ where $\Delta x_{i}$ is one of $\Delta\alpha_{*}$,
$\Delta\delta$, $\Delta\pi$, $\Delta\mu_{\alpha*}$ and $\Delta\mu_{\delta}$,
i.e. given in the catalogue formal errors of the astrometric parameters
and $\sigma_{0}$ is a reference variance (in other words: a 'unit
weight observation uncertainty') for a given star to be cloned. To
calculate $\sigma_{0}$ value one should retrieve three additional
numbers from the stars' record in the catalogue: 'goodness of fit'
- $F2$, 'number of field transits used' - $Ntr$ and 'percentage
of rejected data' - $F1$ and use the following relations:

\begin{align}
dof\ = & \ Ntr(1-F1/100)-5\label{eq:dof-1}\\[2mm]
c\ = & \ 2/(9\cdot dof)\nonumber \\
\chi\ = & \ (1-c+F2\cdot\sqrt{c}){}^{3/2}\cdot dof\label{eq:uwe-1}\\
\sigma_{0}\ = & \ \chi/\sqrt{dof}\nonumber 
\end{align}

The above is a reverse of the procedure used for a 'goodness of fit'
calculation (van Leeuwen, 2013, personal communication). When calculating
the 'degrees of freedom' ($dof$, equation \ref{eq:dof-1}) again
the 5-parameters solution is assumed here. It is worth to note that
formal errors of all five astrometric parameters, namely: $\Delta\alpha_{*}$,
$\Delta\delta$, $\Delta\pi$, $\Delta\mu_{\alpha*}$ and $\Delta\mu_{\delta}$
presented in the catalogue are calculated (van Leeuwen, 2013, personal
communication) by multiplying corresponding diagonal elements from
the covariance matrix by the reference variance $\sigma_{0}$ .

Summarising all the above we obtain the matrix $\mathbf{U}$ in a
form: \begin{small} 
\begin{equation}
\mathbf{U}=\begin{bmatrix}\sigma_{0}u_{1}/\Delta\alpha_{*} & u_{2} & u_{4} & u_{7} & u_{11}\\
0 & \sigma_{0}u_{3}/\Delta\delta & u_{5} & u_{8} & u_{12}\\
0 & 0 & \sigma_{0}u_{6}/\Delta\pi & u_{9} & u_{13}\\
0 & 0 & 0 & \sigma_{0}u_{10}/\Delta\mu_{\alpha*} & u_{14}\\
0 & 0 & 0 & 0 & \sigma_{0}u_{15}/\Delta\mu_{\delta}
\end{bmatrix}\label{eq:matrix U}
\end{equation}
\end{small}

According to equation \ref{eq:z_hip2} we have:

\begin{equation}
\Sigma=(\mathbf{U}^{T}\mathbf{U})^{-1}=\mathbf{U}^{-1}(\mathbf{U}^{T})^{-1}=\mathbf{U}^{-1}(\mathbf{U}^{-1})^{T}=\mathbf{GG}^{T}\label{eq:final_G}
\end{equation}

what shows, that $\mathbf{G}=\mathbf{U}^{-1}$ is suitable for our
purpose of drawing a star clone according to the recipe presented
at the beginning of this section. 

\begin{figure*}
\begin{centering}
\includegraphics[scale=0.7]{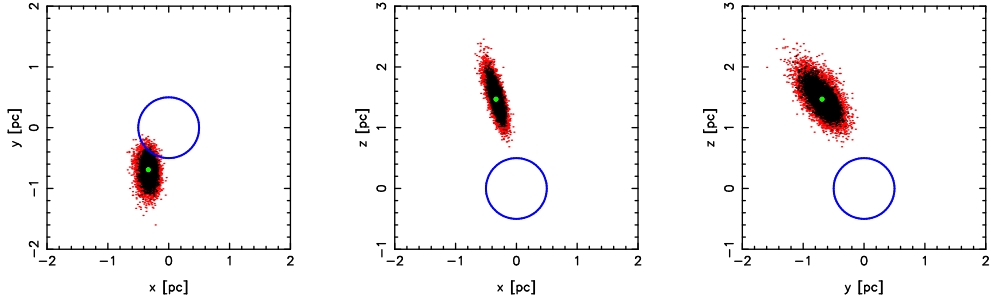} 
\par\end{centering}

\caption{\label{fig:random25240ORG}{\small Differences in a dispersion of
clones of HIP 25240 sampling from six one-dimensional Gaussians (red
dots) and drawing clones from covariance matrix (black dots). All
clones are stopped at the closest proximity epoch. Each green point
mark the nominal result. Blue circle represents here the boundary
of the Oort cloud at 0.5 pc.}}
\end{figure*}

We tested how different methods of drawing clones affect the dispersion
of the proximity position. To do this we repeated our calculation
for all selected stars by drawing all parameters of stellar clone independently.
In most cases distribution was only slightly different in shape, but
in few cases, for example HIP 25240 (see Fig. \ref{fig:random25240ORG})
dispersion was significantly higher than when we used covariance matrix.
This shows that the use of the covariance matrix, apart from being
closer to the data, in some cases can significantly improve the accuracy
of the close stellar approach determination.

\subsection{Galactic motion of stars}

\label{sub:Galactic-motion}

To study the Galactocentric motion of a star we use a numerical integration
of its equations of motion expressed in rectangular coordinates, utilising
the well known, fast and accurate RA15 routine by \citet{everhart-ra15:1985}.
To describe the Galactic gravitational potential we use Model I from
\citet{irrgang_et_al:2013}. Since we are interested in motion of
nearby stars from their current positions backward or forward to the
moment of their closest heliocentric position we deal with rather
small time intervals of order of 10 Myr and small heliocentric distances,
not exceeding 200 pc, with the only one exception of the star HIP~33369
with the current heliocentric distance of 424 pc (such a large distance
makes our results for this star completely unreliable as it is shown
in Section \ref{sec:Results-for-selected}). For such a small solar
vicinity a more sophisticated Galactic potential model, accounting
for non-spherical central bulge or for spiral arms is not necessary,
as was checked for example by \citet{garcia-sanchez:2001} or more
recently by \citet{jimenez_et_al:2011}.

\begin{table}
\caption{\label{tab:Model-I-parameters}Model I parameters from \citet{irrgang_et_al:2013}}

\centering{} %
\begin{tabular}{|c|c|}
\hline 
Parameter  & Value\tabularnewline
\hline 
\hline 
the distance of the Sun from the Galactic centre $R_{\sun}$  & 8400 pc\tabularnewline
Galactic bulge mass $M_{b}$  & 9.51$\times10^{9}\textrm{M}_{\sun}$\tabularnewline
Galactic disk mass $M_{d}$  & 66.4$\times10^{9}\textrm{M}_{\sun}$\tabularnewline
Galactic halo mass $M_{h}$  & 23.7$\times10^{9}\textrm{M}_{\sun}$\tabularnewline
bulge characteristic distance $b_{b}$  & 230 pc\tabularnewline
disk characteristic distance $a_{d}$  & 4220 pc\tabularnewline
disk characteristic distance $b_{d}$  & 292 pc\tabularnewline
halo characteristic distance $a_{h}$  & 2562 pc\tabularnewline
Galactic halo cut-off parameter$\Lambda$  & 200000 pc\tabularnewline
Galactic halo exponent parameter$\gamma$  & 2 (fixed)\tabularnewline
Galactic disk matter density near the Sun $\rho_{o}$  & 0.102 $\textrm{M}_{\sun}/$pc$^{3}$\tabularnewline
Galactic rotational velocity of the LSR $v_{o}$  & 242 km\,s$^{-1}$\tabularnewline
\hline 
\end{tabular}
\end{table}

Gravitational potential $\Phi(r,z)$ considered here is the sum of
a central bulge component $\Phi_{b}(R)$ (spherically symmetric),
an axisymmetric disk $\Phi_{d}(r,z)$ and a massive spherical Galactic
halo $\Phi_{h}(R)$ (dark matter included):

\begin{equation}
\Phi(r,z)=\Phi_{b}(R)+\Phi_{d}(r,z)+\Phi_{h}(R)\label{eq:global_pot}
\end{equation}

where $(r,\varphi,z)$ are Galactocentric cylindrical coordinates
and $R=\sqrt{r^{2}+z^{2}}$ is a Galactocentric spherical radius.
For the Galactic bulge and disk components we use formulae:

\[
\Phi_{b}(R)=-\frac{M_{b}}{\sqrt{R^{2}+b_{b}^{2}}}=\Phi_{b}(x,y,z)=\frac{-M_{b}}{\sqrt{x^{2}+y^{2}+z^{2}+b_{b}^{2}}}
\]

\[
\begin{split}\Phi_{d}(r,z)=-\frac{M_{d}}{\sqrt{r^{2}+\left(a_{d}+\sqrt{z^{2}+b_{d}^{2}}\right)^{2}}}=\\
=\Phi_{d}(x,y,z)=\frac{-M_{d}}{\sqrt{x^{2}+y^{2}+\left(a_{d}+\sqrt{z^{2}+b_{d}^{2}}\right)^{2}}}
\end{split}
\]

and for the Galactic halo we have:

\begin{eqnarray*}
\Phi_{h}(R)=\begin{cases}
\frac{M_{h}}{a_{h}}\left(\frac{1}{\gamma-1}\ln\left(\frac{1+\left(\frac{R}{a_{h}}\right)^{(\gamma-1)}}{1+\left(\frac{\Lambda}{a_{h}}\right)^{(\gamma-1)}}\right)-\frac{\left(\frac{\Lambda}{a_{h}}\right)^{(\gamma-1)}}{1+\left(\frac{\Lambda}{a_{h}}\right)^{(\gamma-1)}}\right) & \textrm{if \ensuremath{R<\Lambda}}\\
-\frac{M_{h}}{R}\frac{\left(\frac{\Lambda}{a_{h}}\right)^{(\gamma-1)}}{1+\left(\frac{\Lambda}{a_{h}}\right)^{(\gamma-1)}} & \textrm{elsewhere.}
\end{cases}
\end{eqnarray*}

what, after adopting $\gamma=2$ and choosing the first equation (we
certainly do not go as far as $R=\Lambda=$200 kpc!) reduces to:

\begin{eqnarray*}
\Phi_{h}(x,y,z) & = & \frac{M_{h}}{a_{h}}\left(\ln\left(\frac{a_{h}+\sqrt{x^{2}+y^{2}+z^{2}}}{a_{h}+\Lambda}\right)-\frac{\Lambda}{a_{h}+\Lambda}\right)
\end{eqnarray*}

The equation of motion of a single point mass (a star) under the potential
described above, expressed in a rectangular Galactic (and Galactocentric)
frame are:

\begin{figure*}
\begin{centering}
\includegraphics[scale=0.7]{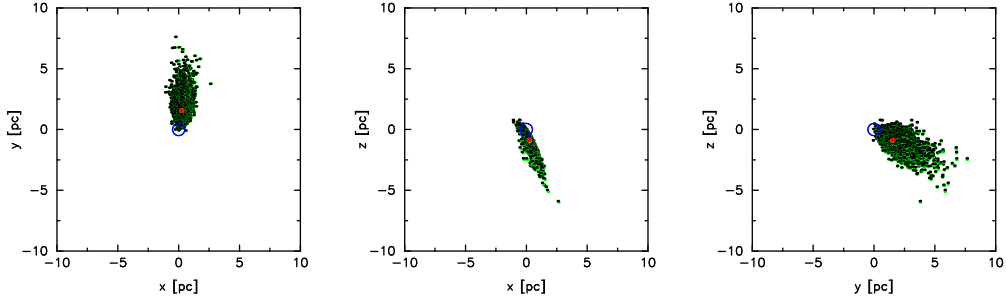} 
\par\end{centering}

\caption{\label{fig:original_frame}{\small Distribution of clones of star
HIP~19946 in the original Galactic heliocentric frame. In this and
all next similar plots black points are the position of clones in
the moment of the closest approach obtained from numerical integration.
Green points are the position of clones calculated with a straight
line motion approximation. Red dot is the position of the nominal
solution from numerical integration, yellow point is the proximity
position for nominal star in a straight line motion model. Blue circle
is the boundary of the Oort cloud at 0.5 pc.}}
\end{figure*}

\[
\begin{aligned}\ddot{x}=-\frac{\partial}{\partial x}\Phi_{b}(x,y,z)-\frac{\partial}{\partial x}\Phi_{d}(x,y,z)-\frac{\partial}{\partial x}\Phi_{h}(x,y,z)\\
\ddot{y}=-\frac{\partial}{\partial y}\Phi_{b}(x,y,z)-\frac{\partial}{\partial y}\Phi_{d}(x,y,z)-\frac{\partial}{\partial y}\Phi_{h}(x,y,z)\\
\ddot{z}=-\frac{\partial}{\partial z}\Phi_{b}(x,y,z)-\frac{\partial}{\partial z}\Phi_{d}(x,y,z)-\frac{\partial}{\partial z}\Phi_{h}(x,y,z)
\end{aligned}
\]

where

\[
\frac{\partial}{\partial x}\Phi_{b}(x,y,z)=\frac{xM_{b}}{\left(x^{2}+y^{2}+z^{2}+b_{b}^{2}\right)^{\frac{3}{2}}}=\frac{xM_{b}}{\left(R^{2}+b_{b}^{2}\right)^{\frac{3}{2}}}
\]

\[
\frac{\partial}{\partial y}\Phi_{b}(x,y,z)=\frac{yM_{b}}{\left(x^{2}+y^{2}+z^{2}+b_{b}^{2}\right)^{\frac{3}{2}}}=\frac{yM_{b}}{\left(R^{2}+b_{b}^{2}\right)^{\frac{3}{2}}}
\]

\[
\frac{\partial}{\partial z}\Phi_{b}(x,y,z)=\frac{zM_{b}}{\left(x^{2}+y^{2}+z^{2}+b_{b}^{2}\right)^{\frac{3}{2}}}=\frac{zM_{b}}{\left(R^{2}+b_{b}^{2}\right)^{\frac{3}{2}}}
\]

\[
\frac{\partial}{\partial x}\Phi_{d}(x,y,z)=\frac{xM_{d}}{\left(x^{2}+y^{2}+\left(a_{d}+\sqrt{z^{2}+b_{d}^{2}}\right)^{2}\right)^{\frac{3}{2}}}
\]

\[
\frac{\partial}{\partial y}\Phi_{d}(x,y,z)=\frac{yM_{d}}{\left(x^{2}+y^{2}+\left(a_{d}+\sqrt{z^{2}+b_{d}^{2}}\right)^{2}\right)^{\frac{3}{2}}}
\]

\[
\frac{\partial}{\partial z}\Phi_{d}(x,y,z)=\frac{zM_{d}\left(a_{d}+\sqrt{z^{2}+b_{d}^{2}}\right)/\left(\sqrt{z^{2}+b_{d}^{2}}\right)}{\left(x^{2}+y^{2}+\left(a_{d}+\sqrt{z^{2}+b_{d}^{2}}\right)^{2}\right)^{\frac{3}{2}}}
\]

\[
\begin{split}\frac{\partial}{\partial x}\Phi_{h}(x,y,z)=\frac{xM_{h}}{a_{h}\sqrt{x^{2}+y^{2}+z^{2}}\left(a_{h}+\sqrt{x^{2}+y^{2}+z^{2}}\right)}=\\
=\frac{xM_{h}}{a_{h}R\left(a_{h}+R\right)}
\end{split}
\]

\[
\begin{split}\frac{\partial}{\partial y}\Phi_{h}(x,y,z)=\frac{yM_{h}}{a_{h}\sqrt{x^{2}+y^{2}+z^{2}}\left(a_{h}+\sqrt{x^{2}+y^{2}+z^{2}}\right)}=\\
=\frac{yM_{h}}{a_{h}R\left(a_{h}+R\right)}
\end{split}
\]

\[
\begin{split}\frac{\partial}{\partial z}\Phi_{h}(x,y,z)=\frac{zM_{h}}{a_{h}\sqrt{x^{2}+y^{2}+z^{2}}\left(a_{h}+\sqrt{x^{2}+y^{2}+z^{2}}\right)}=\\
=\frac{zM_{h}}{a_{h}R\left(a_{h}+R\right)}
\end{split}
\]

Numerical parameters for this model are taken from \citet{irrgang_et_al:2013}
and are summarised in Table \ref{tab:Model-I-parameters}.

Since we always numerically integrate these equations simultaneously
for a star and the Sun (to present final, heliocentric results), we
also account for the gravitational attraction between these two bodies,
as they can approach arbitrarily close. Our test calculation shows
that the greatest influence of taking into account Sun-star gravitational
interaction is for star HIP~26744. This difference in the proximity
distance is 0.00049 pc, less then 0.03 per cent of nominal value for
this star. On this basis, we can state that interactions between Sun
and stars can generally be neglected in such investigations.

\subsection{Principal components analysis}

\label{sub:Principal-components-analysis}

The result of our calculation is always in the form of a swarm of
10\,000 clones stopped at the closest proximity from the Sun. We
have inspected all these swarms of clones in 3D and found, that they
are typically very flat and can be conveniently presented in 2D plots
after applying necessary rotations. To this aim we determine the plane
of the maximum scatter using principal components analysis (PCA).
As can be readed in \citet{Jolliffe:2002} principal component analysis
is one of the oldest techniques of multivariate analysis. Using this
technique, we can find the largest scatter plane for our clones which
is what we would like to present in our plots. Raw results of our calculations
consists of Galactocentric coordinates of each star and the Sun. First
we must go back to heliocentric frame so we need to subtract the calculated
position of the Sun from that of the star.

Then we need to construct a covariance matrix:

\begin{equation}
\Sigma=\begin{pmatrix}cov(\mathbf{x},\mathbf{x}) & cov(\mathbf{x},\mathbf{y}) & cov(\mathbf{x},\mathbf{z})\\
cov(\mathbf{y},\mathbf{x}) & cov(\mathbf{y},\mathbf{y}) & cov(\mathbf{y},\mathbf{z})\\
cov(\mathbf{y},\mathbf{x}) & cov(\mathbf{z},\mathbf{y}) & cov(\mathbf{z},\mathbf{z})
\end{pmatrix}\label{eq:PCA_covMatrix}
\end{equation}

where for example:

\[
cov(\mathbf{x},\mathbf{y})=\frac{\sum_{i=1}^{n}(x_{i}-\overline{x})(y_{i}-\overline{y})}{n-1}=cov(\mathbf{y},\mathbf{x})
\]

and $x_{i}$,$y_{i}$,$z_{i}$ - are coordinates of \textit{i}th clone;
$\overline{x}$,$\overline{y}$,$\overline{z}$ - are respective means
of coordinates of all clones; $n$ - is the number of clones (here
typically 10\,000).

Next we use Jacobi transformation to find eigenvalues and corresponding
eigenvectors. Then we create matrix 3x3 where in first column we place
eigenvector corresponding to the largest eigenvalue and in next columns
eigenvectors corresponding to the remaining eigenvalues. When we have
this matrix with eigenvectors $\mathbf{v}_{i}$ and vector with coordinates
($x$, $y$, $z$) of clone in Galactic heliocentric frame we can
calculate new coordinates ($x'$, $y'$ and $z'$) for all clones
from equation \ref{eq:new_dataPCA}.
\begin{equation}
\begin{pmatrix}x'\\
y'\\
z'
\end{pmatrix}=\begin{pmatrix}\mathbf{v_{1}} & \mathbf{v_{2}} & \mathbf{v_{3}}\end{pmatrix}^{T}\begin{pmatrix}x\\
y\\
z
\end{pmatrix}\label{eq:new_dataPCA}
\end{equation}

\begin{figure*}
\begin{centering}
\includegraphics[scale=0.7]{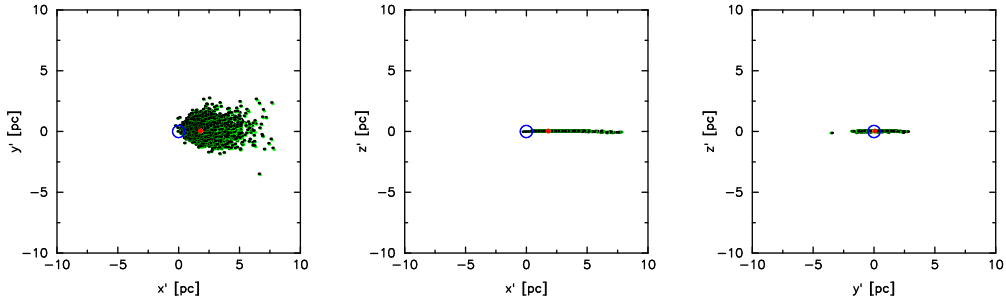} 
\par\end{centering}

\caption{\label{fig:3D_PCA}Distribution of the same swarm of clones of star
HIP~19946 as in Fig. \ref{fig:original_frame} but in $x'y'z'$ heliocentric
frame obtained from the $xyz$ frame after rotations determined using
PCA.}
\end{figure*}

When we look at the results in the original Galactic
heliocentric frame (Fig. \ref{fig:original_frame}) we see a comparable
level of scattering in all planes, but if we express our results with
respect to the frame obtained with PCA (Fig. \ref{fig:3D_PCA}) we
see large difference in scatter level between $x'y'$ plane and the
other two. This clearly reflects the fact that the swarm of clones
stopped at the closest proximity with the Sun is almost flat.

\begin{figure}
\begin{centering}
\includegraphics[scale=0.4]{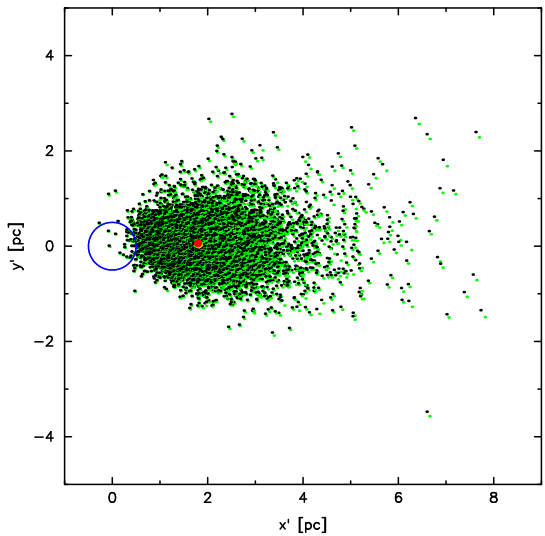} 
\par\end{centering}

\caption{\label{fig:2D_PCA}Distribution of clones HIP~19946 using PCA with
reduction to 2D}
\end{figure}

The final result from the PCA (for plotting purposes) we obtain when
we reduce our data to two dimensions. To reduce dimensions we use
only two eigenvectors corresponding to the two largest eigenvalues
(eq. \ref{eq:reduced_dataPCA}). 
\begin{equation}
\begin{pmatrix}x'\\
y'
\end{pmatrix}=\begin{pmatrix}\mathbf{v_{1}} & \mathbf{v_{2}}\end{pmatrix}^{T}\begin{pmatrix}x\\
y\\
z
\end{pmatrix}\label{eq:reduced_dataPCA}
\end{equation}
 We see that the final plot (Fig. \ref{fig:2D_PCA}) strictly correspond
to the first projection in Fig. \ref{fig:3D_PCA}.

For the star used in the above examples a straight line motion model
is a good approximation for calculating distance of closest approach
to Sun, but in next section we show examples when this approximation
it's not that good. As mentioned earlier a distance calculated with
a rectilinear motion approximation may be significantly different
from a distance obtained from a numerical integration.

\section{Results for selected stars}

\label{sec:Results-for-selected}

\subsection{Selection of stars and overall results}

\label{sub:Selection-of-stars}

In this research we restricted ourselves to the stars from HIP2 catalogue
with known radial velocities. We are interested in the closest stellar
passages so a very small subset of these stars were chosen for our
calculations. The selection procedure was as follows. For each star
in the HIP2 catalogue we checked, if its radial velocity is available
in the XHIP catalogue \citet{anderson_francis:2011} . If it is present,
we calculated the minimal heliocentric distance of this star (in past
or in future) using the rectilinear motion approximation ($D_{l}$).
As a short list of candidates ( 2538 objects) we selected all stars
having this minimal distance $D_{l}$ smaller than 20 pc. We used
such a large threshold value expecting that the exact minimal distance,
obtained from a numerical integration ($D_{min}$), can be substantially
different. In our sample we did not find any star
which $D_{l}>10$ and $D_{min}<2$ pc so the criterion $D_{l}<10$
would be probably sufficient. Than we integrated numerically equations
of motions of every pair, the Sun and each star from the short list,
using full Galactic potential as described in previous section. Our
final list consists of 40 stars with the nominal proximity distance
smaller than 2 pc. In Table \ref{tab:wybrane_40_gwiazd} we present
a complete list of these stars with their common names, nominal proximity
distances and epochs and quality of radial velocity value. In the
last column we describe whether the star is our new findings or its
proximity was noticed and calculated in some earlier papers. In three
cases new stars were independently found by us and by B-J.

\begin{table*}
\caption{\label{tab:wybrane_40_gwiazd}List of stars with nominal proximity
distance smaller than 2 pc. Third column shows quality of radial velocity
quoted from the XHIP catalogue. Results from numerical integration
of nominal data $D_{min}$ and $T_{min}$ are presented in fourth
and fifth column, in next two columns we show for comparison purpose
the results from the rectilinear motion approximation $D_{l}$ and
$T_{l}$.}

\begin{centering}
{\small{} }%
\begin{tabular}{|c|c|c|c|c|c|c|c|}
\hline 
HIP2 ID  & Name  & q\_RV  & $D_{min}$ {[}pc{]}  & $T_{min}$ {[}Myr{]}  & $D_{l}$ {[}pc{]}  & $T_{l}$ {[}Myr{]}  & Earlier publications\tabularnewline
\hline 
\hline 
1392  & HD 1317  & A  & 1.70  & -0.48  & 1.70  & -0.48  & new\tabularnewline
3829  & Van Maanen Star  & D  & 0.95  & -0.02  & 0.95  & -0.02  & a, c, f\tabularnewline
12351  & GJ 1049  & C  & 1.93  & -0.62  & 1.92  & -0.62  & a, c, e, f\tabularnewline
14473  & HD 19376  & A  & 0.35  & -3.73  & 2.37  & -3.78  & new\tabularnewline
14754  & GJ 127.1A  & D  & 1.62  & -0.29  & 1.63  & -0.29  & a, b, e, f\tabularnewline
19946  & BD+03 580  & A  & 1.80  & -0.54  & 1.81  & -0.54  & new\tabularnewline
21539  & CD-33 1835  & D  & 1.92  & -0.14  & 1.92  & -0.14  & new\tabularnewline
23415  & HD 32111  & A  & 1.71  & 5.20  & 3.09  & 5.16  & new\tabularnewline
25001  & HD 34790  & D  & 1.93  & 4.42  & 1.88  & 4.39  & a, c\tabularnewline
25240  & HD 35317  & A  & 1.66  & -0.99  & 1.63  & -0.99  & a, c, f\tabularnewline
26335  & HD 245409  & A  & 1.54  & -0.49  & 1.53  & -0.49  & a, b, c, d, e, f\tabularnewline
26624  & HD 37594  & A  & 1.98  & -1.87  & 1.95  & -1.87  & a, c, e, f\tabularnewline
26744  & HD 37374  & A  & 1.78  & 13.99  & 2.26  & 13.19  & a, c\tabularnewline
27288  & HD 38678  & A  & 1.31  & -0.85  & 1.30  & -0.85  & a, b, c, e, f\tabularnewline
30067  & HD 43947  & A  & 1.78  & -0.66  & 1.78  & -0.66  & a, c, e, f\tabularnewline
30344  & HD 44821  & A  & 1.10  & -1.56  & 1.11  & -1.56  & a, b, e, f\tabularnewline
33369  & V$^{*}$BG Mon  & B  & 0.85  & -5.41  & 3.68  & -5.38  & new\tabularnewline
38228  & HD 63433  & A  & 1.95  & 1.35  & 1.93  & 1.35  & a, b, c, e, f\tabularnewline
38965  & HD 66589  & A  & 1.79  & -1.09  & 1.80  & -1.09  & new, f\tabularnewline
42525  & BD+41 1865  & A  & 0.82  & -0.24  & 0.82  & -0.24  & new, f\tabularnewline
47425  & GJ 358  & C  & 1.87  & -0.06  & 1.87  & -0.06  & a, c, f\tabularnewline
54035  & HD 95735  & A  & 1.44  & 0.02  & 1.44  & 0.02  & a, b, c, d, e, f\tabularnewline
57544  & GJ 445  & A  & 1.06  & 0.05  & 1.06  & 0.05  & a, b, c, d, e, f\tabularnewline
57548  & GJ 447  & A  & 1.92  & 0.07  & 1.92  & 0.07  & a, b, c, d, e, f\tabularnewline
63721  & HD 113447  & A  & 0.12  & 0.13  & 0.12  & 0.13  & new, f\tabularnewline
70890  & Proxima Centauri  & B  & 0.94  & 0.03  & 0.94  & 0.03  & a, b, c, d, e, f\tabularnewline
71681  & $\alpha$ Centauri B  & A  & 0.96  & 0.03  & 0.96  & 0.03  & a, b, c, d, e, f\tabularnewline
71683  & $\alpha$ Centauri A  & A  & 0.98  & 0.03  & 0.98  & 0.03  & a, b, c, e, f\tabularnewline
75311  & BD-02 3986  & A  & 1.62  & 4.84  & 3.55  & 4.92  & a, c\tabularnewline
77910  & HD 142500  & A  & 1.82  & 3.05  & 1.04  & 3.07  & a, c\tabularnewline
84263  & HD 155117  & A  & 1.11  & -6.53  & 2.44  & -6.52  & new\tabularnewline
85661  & HD 158576  & A  & 0.52  & 1.88  & 0.39  & 1.88  & a, c, f\tabularnewline
87052  & HD 161959  & A  & 1.91  & 5.80  & 6.92  & 5.88  & new\tabularnewline
87937  & Bernard Star  & A  & 1.15  & 0.01  & 1.15  & 0.01  & a, b, c, d, e, f\tabularnewline
89825  & GJ 710  & A  & 0.29  & 1.39  & 0.30  & 1.39  & a, b, c, d, e, f\tabularnewline
90112  & HD 168769  & A  & 0.89  & -1.86  & 0.97  & -1.86  & a, c, f\tabularnewline
92403  & GJ 729  & A  & 1.98  & 0.15  & 1.98  & 0.15  & a, b, c, d, e, f\tabularnewline
94512  & HD 179939  & A  & 1.99  & 3.63  & 2.04  & 3.63  & a, c, f\tabularnewline
103738  & HD 199951  & A  & 1.18  & -3.86  & 2.36  & -3.90  & c, f\tabularnewline
110893  & HD 239960  & A  & 1.92  & 0.09  & 1.92  & 0.09  & a, b, c, e, f\tabularnewline
\hline 
\end{tabular}
\par\end{centering}{\small \par}

\medskip{}

a - \citet{jimenez_et_al:2011}, b - \citet{bobylev:2010-1}, c -
\citet{garcia-sanchez:2001}, \\
d - \citet{dyb-kan:1999}, e - \citet{dyb-hab3:2006},
f~-~\citet{bailer-jones:2014}. 
\end{table*}

In columns 5 and 6 of Table \ref{tab:wybrane_40_gwiazd} we present
results (minimal distance $D_{min}$ in parsecs and the corresponding
moment of time $T_{min}$ in Myr from the present epoch) for the nominal
astrometric parameters of each star. The aim of this paper is however
to estimate the accuracy of these results. To achieve this we replaced
each star with 10\,000 of its clones, obtained as described in Section
\ref{sub:Drawing-a-stellar}. Next we numerically integrated motion
of all clones and estimated the distance of the most probable proximity
point and the corresponding epoch with their uncertainties.

The results for all selected stars are presented in Table \ref{tab:wyniki_nowe_dla_40_gwiazd}
and here $D$ (minimal distance) and $T$ (the epoch of the proximity)
are the most probable values obtained from the respective clone distributions.
In some cases they significantly differ from nominal results presented
in Table \ref{tab:wybrane_40_gwiazd}. In the case of moment of time
its calculating procedure was quite simple: $T$ is the mean value
and its uncertainty $\Delta T$ is the half of the symmetric interval
covering 90 per cent of individual clone values. The situation is
much more complicated with the proximity distance estimation, since
their distributions are frequently significantly asymmetric. To present
the most informative result we decided to calculate three mean coordinates
of the most probable proximity point in space (calculating simple
means in each coordinate, i.e. $\overline{x'}$, $\overline{y'}$
and $\overline{z'}$ ) and then to calculate its distance from the
Sun:
\begin{equation}
D=\sqrt{\overline{x'}^{2}+\overline{y'}^{2}+\overline{z'}^{2}}\label{eq:meanR}
\end{equation}

To estimate its uncertainty we measure the radius of a sphere around
the proximity point in which 90 per cent of individual clone proximities
are placed and this value is presented in Table \ref{tab:wyniki_nowe_dla_40_gwiazd}
as $\Delta D$.

\begin{table*}
\caption{\label{tab:wyniki_nowe_dla_40_gwiazd}Estimated minimum distance R
from the Sun and corresponding moment of time T for 40 stars that
nominally can came closer than 2 pc. A dash marks cases where we cannot
obtain reasonable values.}

\centering{}
\begin{tabular}{|r|r|r|r|r|}
\hline 
HIP ID & $D$ {[}pc{]} & $\Delta D$ {[}pc{]} & $T$ {[}Myr{]} & $\Delta T$ {[}Myr{]}\tabularnewline
\hline 
1392~~~ & 1.70 & 0.16 & -0.48 & 0.02\tabularnewline
3829 {*} & 0.95 & 0.05 & -0.02 & <0.01\tabularnewline
12351~~~ & 2.26 & 1.47 & -0.71 & 0.45\tabularnewline
14473 ? & 0.22 & 7.84 & -3.78 & 0.74\tabularnewline
14754~~~ & 1.64 & 0.27 & -0.29 & 0.04\tabularnewline
19946~~~ & 1.91 & 1.32 & -0.55 & 0.13\tabularnewline
21539 {*} & 1.97 & 0.49 & -0.14 & 0.03\tabularnewline
23415 ? & 1.65 & 7.82 & 5.35 & 1.43\tabularnewline
25001~~~ & 2.07 & 1.54 & 4.67 & 1.70\tabularnewline
25240~~~ & 1.66 & 0.30 & -0.99 & 0.05\tabularnewline
26335~~~ & 1.54 & 0.09 & -0.49 & 0.01\tabularnewline
26624~~~ & 1.98 & 0.25 & -1.87 & 0.08\tabularnewline
26744 ? & 1.71 & 4.21 & 14.12 & 2.30\tabularnewline
27288~~~ & 1.31 & 0.10 & -0.85 & 0.06\tabularnewline
30067~~~ & 1.78 & 0.11 & -0.66 & 0.01\tabularnewline
30344~~~ & 1.11 & 0.23 & -1.57 & 0.11\tabularnewline
33369 ? & 51.00 & -- & -4.96 & --\tabularnewline
38228~~~ & 1.95 & 0.16 & 1.35 & 0.03\tabularnewline
38965~~~ & 1.87 & 1.16 & -1.10 & 0.23\tabularnewline
42525~~~ & 0.99 & 0.96 & -0.25 & 0.09\tabularnewline
\hline 
\end{tabular}
\begin{tabular}{|r|r|r|r|r|}
\hline 
HIP ID & $D$ {[}pc{]} & $\Delta D$ {[}pc{]} & $T$ {[}Myr{]} & $\Delta T$ {[}Myr{]}\tabularnewline
\hline 
47425 {*} & 1.90 & 0.45 & -0.06 & 0.01\tabularnewline
54035 {*} & 1.44 & 0.01 & 0.02 & <0.01\tabularnewline
57544 {*} & 1.06 & 0.03 & 0.05 & <0.01\tabularnewline
57548 {*} & 1.92 & 0.04 & 0.07 & <0.01\tabularnewline
63721 ? & 0.13 & 0.04 & 0.14 & 0.03\tabularnewline
70890 {*} & 0.94 & 0.03 & 0.03 & <0.01\tabularnewline
71681 {*} & 0.96 & 0.09 & 0.03 & <0.01\tabularnewline
71683 {*} & 0.98 & 0.04 & 0.03 & <0.01\tabularnewline
75311 ? & 5.71 & -- & 5.23 & --\tabularnewline
77910~~~ & 1.94 & 1.59 & 3.15 & 0.90\tabularnewline
84263 ? & 1.20 & 4.64 & -6.59 & 1.07\tabularnewline
85661~~~ & 0.52 & 0.62 & 1.88 & 0.11\tabularnewline
87052 ? & 1.81 & 5.87 & 5.90 & 1.27\tabularnewline
87937 {*} & 1.15 & 0.01 & 0.01 & <0.01\tabularnewline
89825~~~ & 0.29 & 0.35 & 1.39 & 0.08\tabularnewline
90112~~~ & 0.91 & 1.31 & -1.87 & 0.20\tabularnewline
92403 {*} & 1.98 & 0.04 & 0.15 & <0.01\tabularnewline
94512~~~ & 2.06 & 2.33 & 3.65 & 0.46\tabularnewline
103738~~~ & 1.18 & 0.69 & -3.87 & 0.30\tabularnewline
110893 {*} & 1.92 & 0.04 & 0.09 & <0.01\tabularnewline
\hline 
\end{tabular}
\end{table*}

From the inspection of the data in Table \ref{tab:wyniki_nowe_dla_40_gwiazd}
several observations can be made. First, there are two stars, HIP~33369
and HIP~75311 for which our results are completely unreliable. First
star is at the exceptionally large distance at the present epoch (424
pc) and any uncertainties in its position and/or velocity significantly
amplifies when numerically integrating its motion back to the solar
proximity. While its nominal proximity distance is only 0.85 pc we,
basing on the contemporary astrometric data for this star, cannot
say anything reliable about its real minimal distance from the Sun
5 Myr ago.

For the second star, HIP~75311, the main source of the unacceptable
uncertainty in the proximity distance is the large formal error of
its proper motion. Both components of the proper motion of this star
have uncertainties on the level of 300 per cent in the HIP2 catalogue.
The nominal proximity distance is 1.62 pc, the most probable value
equals 5.71 and both should be treated as highly unreliable. Both
these stars are marked with '?' in Table \ref{tab:wyniki_nowe_dla_40_gwiazd}
as well as five more stars due to their large proximity position errors.
There is also one additional star marked in the same way despite of
its small error, namely HIP~63721. The reason \textbf{for this} is that we are aware
of its highly unreliable parallax in HIP2. Instead of $\pi=217$ mas
included in HIP2 other sources present parallax value well beyond
5 mas (see for example \citet{fabricius-makarov:2000}). HIP2 value
is also discarded in XHIP catalogue. 

\begin{figure}
\centering{}\includegraphics[scale=0.4]{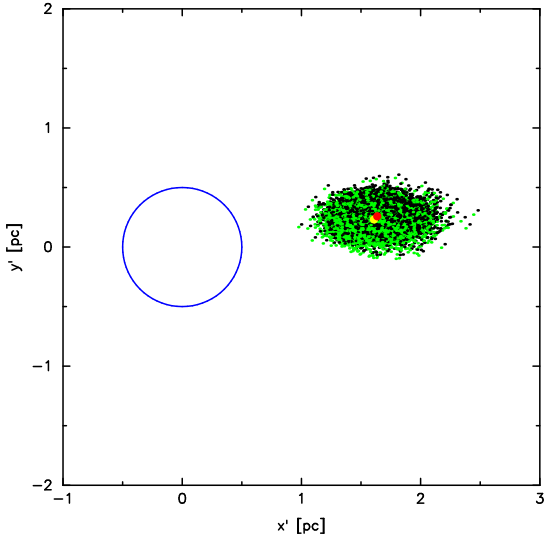}\caption{\label{fig:PCA_HIP25240_example_accu}HIP~25240 - the example of
a compact swarm of clones resulting from a current distance of this
star which is relatively close to the minimum value (nominally 1 Myr
ago).}
\end{figure}

There are also 12 stars (marked with '{*}' just after their HIP numbers
in Table \ref{tab:wyniki_nowe_dla_40_gwiazd}) in our final list which
are now close to the Sun and simultaneously close to their proximity
epoch. For these stars a proximity uncertainty is practically equal
to their current astrometric position and velocity uncertainties and
therefore small or very small. In our calculations we integrated them
only over extremely short time intervals and we did not observe any
uncertainty amplification. The example of such a result is presented
in Fig.\ref{fig:PCA_HIP25240_example_accu}. Any further discussion
of the accuracy of our results for these stars seems unnecessary.

\subsection{Examples}

\label{sub:Examples}

The remaining 20 stars present a wide variety of proximity distance
accuracy estimations. We will discuss several representative examples
in following subsections.

\subsubsection{HIP 89825 ( Gliese 710)}

\label{sub:HIP-89825}

\begin{figure}
\begin{centering}
\includegraphics[scale=0.4]{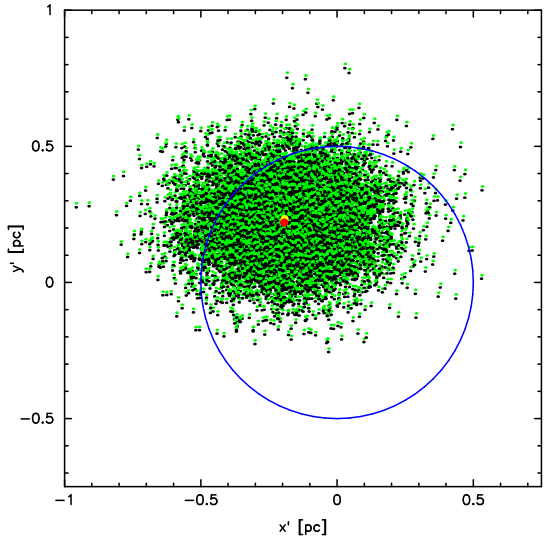} 
\par\end{centering}

\caption{\label{fig:PCA_Gliese710}Distribution of clones Gliese~710 using
RV=-13.80 km\,s$^{-1}$}
\end{figure}

This star is a well known future visitor in the solar neighbourhood,
see for example \citet{mullari-o:1996,dyb-kan:1999,garcia-sanchez:2001,dyb-hab3:2006}.
According to the latest estimation of its astrometric parameters presented
in HIP2 catalogue and using $v_{r}=-13.8$ km\,s$^{-1}$ (\citet{Gontcharov:2006},
used also in XHIP) we obtained the proximity distance (both nominal
and mean) as small as 0.29 pc. As it is clearly depicted in Fig.\ref{fig:PCA_Gliese710}
the straight linear motion approximation works very well in this case.
The swarm of clones of this star is however significantly dispersed,
mainly due to large uncertainties of proper motions of Gliese 710.
If one had used an older value of its radial velocity: -23 km\,s$^{-1}$
\citep{wilson:1953}, later confirmed for example in the Palomar/MSU
survey \citep{PalomarMSU:1995} , the minimal distance would reduce
even further, down to 0.17 pc and almost all clones of this star would
be situated closer than 0.5 pc from the Sun i.e., the widely accepted
radius of the outer Oort cloud.

\subsubsection{HIP 14473 (HD 19376)}

\label{sub:HIP-14473}

\begin{figure}
\begin{centering}
\includegraphics[scale=0.4]{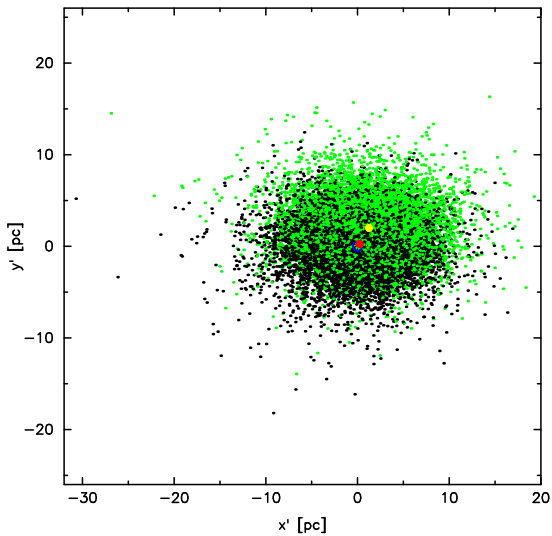} 
\par\end{centering}

\caption{\label{fig:PCA_HIP14473}Distribution of clones HIP~14473}
\end{figure}

This star is our new finding and the only one in our list for which
an estimated proximity distance in the past (3.78 Myr ago) is significantly
smaller than the adopted radius of the Oort cloud. We estimated this
minimal distance to be $D=$0.22 pc with the proximity position uncertainty
$\Delta D=$7.84 pc (see Fig.\ref{fig:PCA_HIP14473}). The reason of such a large error of our estimation
is a large uncertainty of proper motion of this star. If one obtain
much more accurate value (for example from Gaia mission) this star
should be considered as a serious candidate for the stellar perturber
of cometary motion in the past.

\subsubsection{HIP 103738 (HD 199951)}

\label{sub:HIP-103738}

\begin{figure}
\centering{}\includegraphics[scale=0.4]{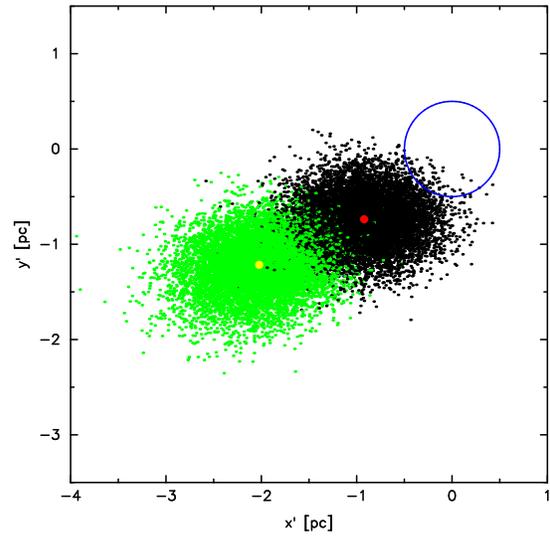}\caption{\label{fig:PCA_HIP103738} Distribution of clones of HIP~103738}
\end{figure}

Here we have a good example of the significant difference between
the exact numerical integration in the Galactic field and the rectilinear
approximation (see Fig.\ref{fig:PCA_HIP103738}). Using the numerical
integration we obtained the minimal distance smaller by a factor of
two. This star is now relatively distant ( about 70 pc) but all its
positional and kinematic parameters are known with such a good accuracy
that even long term numerical integration (for almost 4 Myr) do not
downgrade the precision of the result significantly.

\subsubsection{HIP 87052 (HD 161959)}

\label{sub:HIP-87052}

\begin{figure}
\begin{centering}
\includegraphics[scale=0.4]{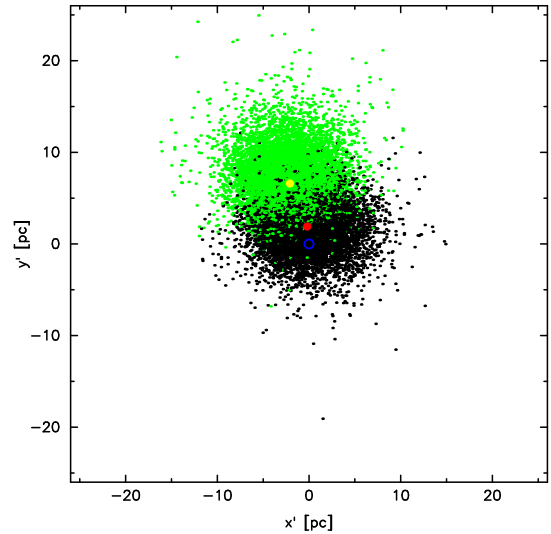} 
\par\end{centering}

\caption{\label{fig:PCA_HIP87052} Distribution of clones HIP~87052}
\end{figure}

\begin{figure}
\begin{centering}
\includegraphics[scale=0.4]{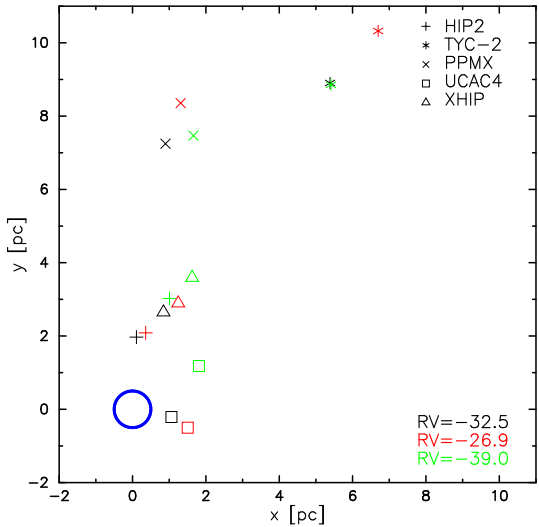}
\par\end{centering}

\caption{\label{fig:kat_87052}Proximity point positions of HIP~87052 obtained
from different catalogue data. This plot is a projection on the plane
perpendicular to present heliocentric direction to this star. Different
symbols correspond to different source catalogues and different colours
correspond to three different radial velocity values used in this
plot.}
\end{figure}

This star is another one among 10 stars recognised in this paper for
the first time as possibly visiting the solar vicinity. The reason
for this novelty is that the straight line motion approximation (green
points in Fig. \ref{fig:PCA_HIP87052}) gives the proximity distance
for this star as large as almost 7 pc! The distance to the
most probable proximity point from numerical integration obtained
by us is only 1.81 pc but its dispersion is large so we estimated
rather large error for this value, namely $\pm$5.87 pc. This is the
consequence of a large uncertainty of proper motion for this star.
This is also present when we compare nominal proximity points obtained
on the basis of different astrometric catalogues, as it is depicted
in Fig. \ref{fig:kat_87052}.

\subsubsection{HIP 84263 (HD 155117)}

\label{sub:HIP-84263}

\begin{figure}
\begin{centering}
\includegraphics[scale=0.4]{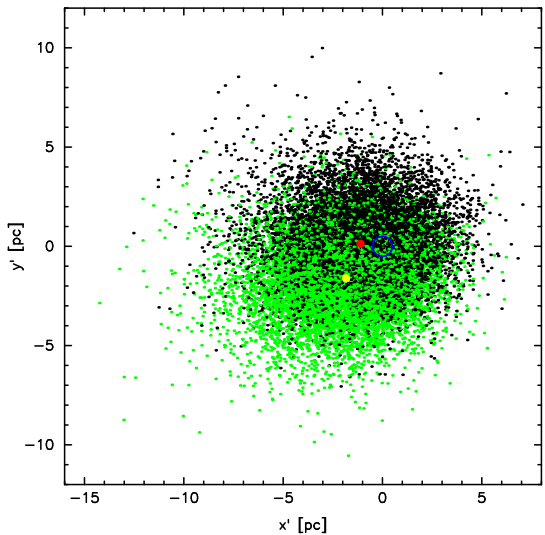} 
\par\end{centering}

\caption{\label{fig:PCA_HIP84263}Distribution of clones of HIP~84263}
\end{figure}

In this case the minimal distance obtained from a numerical integration
reduces to 1.20$\pm$4.64 pc comparing with 2.44 from a rectilinear
model. This star has very small proper motion with rather big relative
errors which together with the long numerically integrated time interval
(over 6 Myr) gives a widely spread swarm of clones, with significant
number located inside the assumed Oort cloud sphere (see Fig. \ref{fig:PCA_HIP84263}).

\subsubsection{HIP 77910 (HD 142500)}

\label{sub:HIP-77910}

\begin{figure}
\begin{centering}
\includegraphics[scale=0.4]{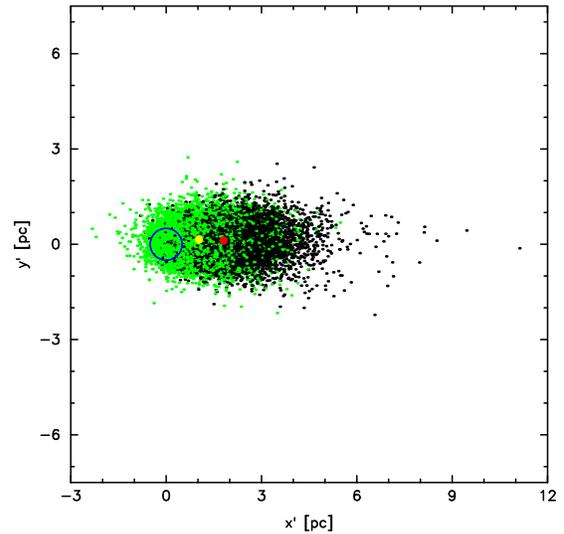} 
\par\end{centering}

\caption{\label{fig:PCA_HIP77910}Distribution of clones of HIP~77910}
\end{figure}

Using numerical integration with full Galactic potential we almost
double the minimal distance of HIP~77910 from the Sun, obtaining
$D=1.94\pm1.59$ pc (black points in Fig. \ref{fig:PCA_HIP77910}).
This makes its gravitational influence on Oort cloud comets weaker
but since its estimated mass is 2 M$_{\sun}$ it should still be considered
as a potential stellar perturber.

\subsubsection{HIP 57791 (HD 102928)}

\label{sub:HIP-57791}

\begin{figure}
\begin{centering}
\includegraphics[scale=0.4]{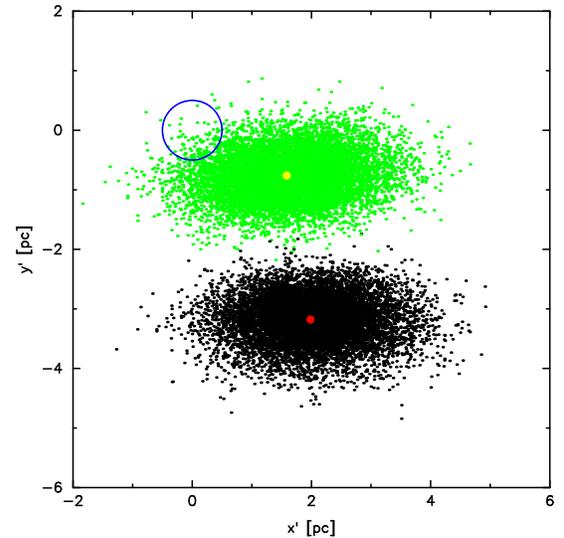} 
\par\end{centering}

\caption{\label{fig:PCA_HIP57791}Distribution of clones HIP~57791}
\end{figure}

HIP~57791 is an extra example (outside our star list but for example
included by \citealp{garcia-sanchez:1999}) of completely different
results from the rectilinear approximation and numerical integration.
In this case a full model moves out the proximity point from 1.8 pc
to 3.7 pc from the Sun (see Fig. \ref{fig:PCA_HIP57791}) what makes
this star rather improbable stellar perturber for the Oort cloud comets.

\section{'Last minute' discussion}

\label{sec:Just-in-the}Just in the moment we had our paper ready
for submission we learned about a very recent paper by \citet{bailer-jones:2014}.
He used almost identical data and methods and obtained results similar
to ours in many respects. However there are also several differences
in our approach and as a consequence his results are in many cases
significantly different so it is worth to study in detail each individual
star case. 

In his Table 3 B-J\textcolor{red}{{} }describes 65 close (closer than
2 pc from the Sun) stellar passages. However, for many stars he included
several variants (based on different stellar data) so in fact he presented
stellar proximities for 42 individual stars. This is very close to
40 stars in our Tables \ref{tab:wybrane_40_gwiazd} and \ref{tab:wyniki_nowe_dla_40_gwiazd}
but surprisingly we have only 27 stars in common. There are two main
reasons for this discrepancy. First, B-J presents his results in significantly
different manner. He calculate simple mean of all proximity distances
for all clones with 90 per cent confidence intervals and in many cases this method somewhat overestimates
the distance we should expect. The heliocentric distance is obviously
positive so its distribution is often significantly asymmetric. Moreover
mean value is rather sensitive to outliers and due to the limitation
to positive values almost all outliers are placed at one side of the
mean. For this reason we decided to present the distance to the most
probable proximity position, as it is described in detail in Sec.\ref{sub:Selection-of-stars}. Since B-J selected stars for his published Table 3 
using only mean distances several objects were shifted out to his Internet large table.
The most spectacular example is the star HIP~14473. As it is shown
in Fig.\ref{fig:PCA_HIP14473} spatial positions of proximity points
of individual clones are widely spread. B-J calculated the mean of
all this distances and report the result (in his extended, Internet
table) to be 6 pc. In our approach we report the proximity distance
as 0.22$\pm$7.84 pc (of course the second number is the uncertainty
in proximity point position, not the formal error in the proximity
distance which must be positive).

The second reason for several discrepancies between B-J and our results
is his usage of few questionable radial velocities from RAVE catalogue.  The great majority of velocities taken by B-J from RAVE catalogue seems to be obviously erroneous, in some cases greater than the escape velocity from our
Galaxy. B-J also commented all this cases as uncertain. In our investigation we did not use this velocities. 

There are also several other differences in the methods (e.g. the Galaxy
potential model) and result presentation, including the main purpose
of the work: we aimed at studying and presenting the accuracy of proximity
calculations while B-J concentrate on completing the list of closest
stellar approaches. He also discuss in detail many problematic cases which helped us to reveal the source of some discrepancies, e.g  for HIP 85605.

As it concerns the influence of the Galaxy potential model, B-J used
the Dauphole and Colin model \citeyearpar{dauphole-colin:1995,dauphole:1996}
and states: ``While better models may now exist, the results are
not very sensitive to the exact choice''. The influence of different
Galaxy models for close stellar approaches to the Sun was recently
discussed by \citet{jimenez_et_al:2011} and their conclusion is that
for more distant stars the results are sensitive to the choice of
the model. They present a large table comparing results for different
models (their tab.3) but erroneously instead of presenting their own
results they copied results form \citet{garcia-sanchez:2001}. Since
we used a very recent model by \citet{irrgang_et_al:2013} we are
interested in such a comparison.

B-J kindly made available for us his nominal results (Bailer-Jones,
2015, personal communication). Thanks to that we were able to study
the difference between stellar results for two models in use (including
slightly different initial Sun position and velocity and marginal
difference in the Galactic frame orientation). We found moderate differences
in the nominal closest stellar approach distances for several objects.
In Table \ref{tab:Differences-du-to} we present selected examples
of such differences (for the purpose of this comparison we used exactly
the same stellar initial data as B-J).

\begin{table}

\caption{\label{tab:Differences-du-to}The examples of the minimal star-Sun
distances (nominal) calculated with different Galaxy potential model.
$D_{min}^{1}$ is calculated by B-J (Bailer-Jones, 2015, personal
communication) with the Dauphole and Colin model \citeyearpar{dauphole-colin:1995,dauphole:1996}.
$D_{min}^{2}$ is calculated by us with the model used in this paper
\citep{irrgang_et_al:2013}. Object names consist of the HIP number
augmented with a letter indicating the source of a radial velocity,
detailed explanation can be found in B-J.}

\begin{centering}
\begin{tabular}{ccc}
\hline 
object & $D_{min}^{1}${[}pc{]}  & $D_{min}^{2}$ {[}pc{]}\tabularnewline
\hline 
14473g & 0.072 & 0.375\tabularnewline
14473p & 0.095 & 0.353\tabularnewline
75159r & 0.347 & 0.251\tabularnewline
23415p & 1.486 & 1.715\tabularnewline
84263x & 1.419 & 0.802\tabularnewline
84263p & 0.723 & 1.112\tabularnewline
84263g & 0.731 & 1.117\tabularnewline
87052p & 0.394 & 1.914\tabularnewline
87052x & 1.409 & 2.736\tabularnewline
91012r & 0.446 & 0.350\tabularnewline
\hline 
\end{tabular}
\par\end{centering}

\end{table}

In the following sections we discuss in detail all discrepancies we
found in the stellar close approaches lists of B-J and ours.

\subsection{14 stars present in B-J but omitted here}

There are two groups of stars from B-J list missing in our paper.
First is the result of using RAVE radial velocities (not used by us)
and the second comes from the utilisation of astrometric parameters
(mainly proper motions) from the XHIP catalogue, where HIP2 astrometry
was replaced by some other, frequently taken from HIP1 or Tycho-2. 

From the first group (10 stars), in our opinion, only star HIP~87784
can be treated as having close approach to the Sun (assuming that
the RAVE radial velocity of -66.30$\pm$3.60 km\,s$^{-1}$ found
by B-J will be confirmed). Also HIP~75159 might be considered as
real (but highly uncertain) solar neighbourhood visitor. The rest
(HIP~23311, HIP~41312, HIP~53911, HIP~55606, HIP~91012, HIP~100280,
HIP~104256 and HIP~104644) are results of using unacceptable radial
velocities.

Coming to the second group: 
\begin{itemize}
\item HIP~34617 is the double system and proper motions in HIP-2 are not
necessarily the best, so probably B-J result, using Tycho-2 values,
is better. 
\item HIP~85605 is also a double system and erroneous parallax for this
star exists in Hipparcos, as pointed out by SIMBAD \citep{SIMBAD-article:2000}
database comment (``Parallax and proper motion are not compatible
with CCDM J17296+2439A; the large proper motion and parallax of this
star in Hipparcos ($\pi$=202mas, $\mu$=362mas/yr) is most likely
an artefact''). The similar opinion is expressed in \citet{fabricius-makarov:2000}.
\item HIP~86961 and HIP~86963 are members of the triple system and according
to \citet{henry:2006} both these stars have significantly erroneous
parallaxes in Hipparcos catalogues.
\end{itemize}

\subsection{12 stars present in this paper but omitted in B-J}

Apart from some differences in Galaxy potential model described above
the main reason for some our stars missing in B-J list is presenting
the mean value of the closest Sun-star distance. In many cases this
results in larger values than presented by us, which includes stars
HIP~1392, HIP~14473, HIP~19946, HIP~23415, HIP~25001, HIP~26744,
HIP~77910, HIP~84263 and HIP~87052. For all these 9 stars the nominal
proximity distance calculated by B-J is below the 2 pc threshold but
his mean distance is larger and he excluded them from his published
list.

There are three more stars in our list that are omitted by B-J. First
is star HIP~21539 for which nominal proximity distance for data taken
from XHIP catalogue is 1.92 pc but uncertainty of its radial velocity
is undetermined (see Sec. \ref{sub:Drawing-a-stellar}) and probably
for that reason B-J removed this star from his sample. Next two stars
are HIP~33369, our nominal result equals 0.85, and HIP~75311 - HIP2
astrometry combined with the radial velocity from Pulkovo catalogue
gives nominal proximity distance of 1.62 pc. These two stars have
enormously large uncertainty of the proximity point position so their
mean distance values are larger then 10 pc (and therefore omitted
by B-J), but their nominal solutions are below the 2 pc limit.

\subsection{Other difficult cases}

As we have all nominal solutions kindly made available by B-J (also
those corresponding to his large, Internet table) we carefully compared
his results with ours. Few problematic cases we found for stars with
the nominal proximity distance smaller than 2 pc:
\begin{itemize}
\item For HIP~1647 B-J found a radial velocity of -86.8 $\pm$ 28.8 km\,s$^{-1}$
in RAVE catalogue and obtained the closest nominal distance of 1.979
pc. While in our model this distance is even smaller, 1.881 pc we
think this result should be discarded. The reason is that HIP~1647
have a well established (since 1928) radial velocity of +12 km\,s$^{-1}$,
i.e. in opposite direction.
\item For HIP~10332 B-J have obtained a minimal, nominal distance of 1.793
pc basing on XHIP data. While XHIP authors treat this star astrometry
as problematic and incorporate (a little bit inconsistently) its parallax
from HIP1 and its proper motion from Tycho-2, the problem seems to
be more serious. In \citet{luck:2011} the authors obtained the heliocentric
distance for this star to be over 4 kpc. HIP~10332 (V$^{*}$UX Per)
is a classical $\delta$ Cepheid type star with the period over one
year. There was well known problem in Hipparcos mission with obtaining
correct parallaxes of classical cepheides and this star is most probably
one of the problematic Hipparcos results.
\item B-J used also XHIP data for HIP~24670 obtaining the nominal proximity
of 1.729 pc. Proper motions of this star are very small with large
uncertainties in all catalogues, which connected with its large current
distance (over 160 pc according to HIP2) makes the uncertainty of
the proximity distance very large. Mean value from B-J equals 5.3
pc.
\item HIP~32475 according to B-J has the close approach nominal distance
of 1.725 pc basing on XHIP data, where TYCHO-2 proper motions are
included. This star is in fact a close binary (CCDM J06467+0822AB),
unresolved in Hipparcos and Tycho catalogues. If we treat TYCHO-2
proper motions as superior to HIP2 this B-J result is fully valid.
\item For HIP~75807 B-J found a radial velocity of -37.40$\pm$1.20 in
RAVE catalogue which, combined with HIP2 astrometry gives a nominal
proximity distance of 1.461 pc. The problem is that the HIP2 astrometry
for this star (especially its parallax) is evidently in error. In
the same RAVE catalogue one can find the spectrophotometric parallax
for this star equal to 2.55 mas, which is highly consistent with HIP1
result, further corrected by \citet{fabricius-makarov:2000}.
\end{itemize}

\section{Conclusions}

\label{sec:Conclusions}

The main purpose of this investigation was to study the accuracy of
determination of the proximity distance and epoch for stars that can
come ( in past or in future) close to the Sun and act as stellar perturbers
of the long period comet motion. Our study is based on the best available
data in the pre-Gaia era, mainly from the HIP2 \citep{vanleeuwen:2011}
and XHIP \citep{anderson_francis:2011} catalogues. We used a numerical
integration in the rectilinear, Galactocentric coordinates taking
into account the full gravitational potential of the Galaxy in its
modern shape \citep{irrgang_et_al:2013}. For the main purpose (the
accuracy assessment) we utilised covariance matrices of astrometric
data which is included in the HIP2 catalogue. Each star was substituted
by a swarm of 10\,000 of its clones according to the respective covariance
matrix and radial velocity dispersion. Than we used Principal Component
Analysis to present the distribution of the swarm at the proximity
epoch.

Limiting our self to the nominal proximity distance of 2 pc we formulated
a list of 40 stellar close visitors and among them 10 stars were our
independent new findings (three of them were independently pointed
out by B-J). We analysed each case in detail and concluded that for
more than 50 per cent of stars in our list the accuracy is good or
very good because these stars are now close to their proximity epoch.
We showed several examples of moderate accuracy for more distant (in
space and in time) stars.

We also showed that the linear approximation method can lead to large
errors in many cases, especially for more distant stars, where the
curvature in their motion induced by a Galaxy gravity cannot be ignored. 

We also concluded that for two stars (HIP~33369 and HIP~75311) the
extremely large formal errors makes our proximity distance results
completely unreliable. We suspect that the result for HIP~63721 is
also erroneous due to its probably false parallax in the HIP2 catalogue.
We also marked next five stars ( HIP~14473, HIP~23415, HIP~26744,
HIP~84263 and HIP~87052) as having unacceptably large estimated
errors. The first of them might be an important perturber of the long
period comets motion since its nominal proximity distance was as small
as 0.22 pc about 3.78 Myr ago. The problem is that the proximity position
uncertainty ($\Delta D=7.84$) is unacceptably large due to the astrometry
uncertainties. 

We are convinced that the incoming advance in stellar astrometry caused
by Gaia mission can significantly improve the accuracy of nearby stars
passages close to the Sun. The most important part is the precision
of proper motions and parallaxes, augmented with (mainly ground based)
precise radial velocity measurements.

\section*{Acknowledgements}

We would like to express our gratitude to Coryn Bailer-Jones for making
available his nominal results and detailed discussion on the results.
We also wish to thank Floor van Leeuwen for the detailed description
how to obtain full covariance matrices from the HIP2 catalogue records.
Many constructive coments from Małgorzata Królikowska allowed us to
improve this manuscript. This research has made use of the SIMBAD
database, operated at CDS, Strasbourg and of NASA's Astrophysics Data
System.

\bibliographystyle{mn2e}
\bibliography{moja23}

\end{document}